# Towards 3D AI Hardware: Fine-Grain Hardware Characterization of 3D Stacks for Heterogeneous System Integration & AI Systems

Eren Kurshan
ekurshan@princeton.edu
Princeton University

Paul D. Franzon
paulf@ncsu.edu
North Carolina State University

## Abstract

3D integration offers key advantages in improving system performance and efficiency for the End-of-Scaling era. It enables the incorporation of heterogeneous system components and disparate technologies, eliminates off-chip communication constraints, reduces on-chip latency and total power dissipation. Moreover, AIs demand for increased computational power, larger GPU cache capacity, energy efficiency and low power custom AI hardware integration all serve as drivers for 3D integration. Although 3D advantages such as enhanced interconnectivity and increased performance have been demonstrated through numerous technology sites, heterogeneous 3D system design raises numerous unanswered questions. Among the primary challenges are the temperature and lifetime reliability issues caused by the complex interaction patterns among system components. Such interactions are harder to model with current modeling tools and require detailed hardware characterization. This study presents the latest drivers for 3D integration and the resulting need for hardware emulation frameworks. It then presents a design to profile power, temperature, noise, inter-layer bandwidth and lifetime reliability characterization that can emulate a wide range of stacking alternatives. This framework allows for controlling activity levels at the macro-level, along with customized sensor infrastructure to characterize heat propagation, inter-layer noise, power delivery, reliability and interconnectivity as well as the interactions among critical design objectives.

## 1. Introduction

The demand for increased computational power has significantly influenced system-level design over the past decade. These trends have resulted in increased number of cores on-chip, larger chip footprints, and higher power dissipation. The end of Dennard Scaling and Moore's Law posed serious challenges in driving continuous performance growth in computing systems [37]. Not only is each new technology generation significantly more expensive to manufacture, achieving performance improvement in the post-scaling era requires innovation across the full design stack. Recently, emerging technologies and system-level optimization have gained significant attention as potential means to sustain performance growth trends while improving energy efficiency and enabling modularity at a low cost. 3D integration, which lies at the intersection of these key areas, has been considered a promising alternative to scaling. 3D integration refers to a broad range of technologies such as die-to-die, die-to-wafer and wafer-to-wafer bonding [39]. It provides key advantages by enabling the integration of more transistors in the same chip footprint, heterogeneous integration of disparate manufacturing technologies, increasing the on-chip interconnectivity, reducing power dissipation and ultimately providing improved system performance at a lower cost. [19].

## 2. 3D Drivers in Computing Systems

### 2.1. New Scaling Laws in the End-of-Scaling Era

*2.1.1* **AIs Increasing Computational Demands:** In recent years, AI has made rapid progress in a number of narrow AI tasks and has been making strides towards artificial general intelligence (AGI). According to some reports, since the introduction of Kaplan scaling, which asserts that increasing the number of parameters of an AI solution provides higher performance, the amount of computation and energy required to train each AI solution has been doubling every 2 months [38]. This represents a sharp increase compared to the doubling every 3.4 months a few years ago. As more players and users enter the market and AI's parameter scaling persists, the demand for computation will continue reaching unprecedented levels.

*2.1.2* **AIs Energy Challenges:** In addition to its unparalleled computing demands, training energy and run-time power dissipation have been increasing rapidly. This trend has become so powerful that it became the primary limitation for current AI systems. Energy efficiency challenges first came to the spotlight in the early days of commercial AI solutions - when reports of AlphaGo consuming orders of magnitude higher run-time power than its human competitors emerged (around 170kW vs 20W)[16]. Similarly, GPT-4's daily energy consumption is estimated to be around 260.42 MWh [17]. According to recent estimates, a *ChatGPT* query requires nearly *10 times* as much electricity to process as a *Google search*. GPU training energy budgets have become primary limitations of AI development. This caused a new wave of interest in low power and energy efficient architectures, custom low power AI hardware and 3D system integration.

*2.1.3* **Neuromorphic Integration with Digital Systems:** Neuromorphic computing provides higher performance and orders of magnitude lower run-time power dissipation for AI. Though there is significant variance among neuromorphic chips, spike based computation is a different paradigm than digital computing. In neuromorphic chips, voltage spikes often do not carry information, but the intervals between spikes carry the signal making it harder to integrate with standard digital systems. In the past years, 3D integration of neuromorphic chips with standard digital processors and interface components is seen as a promising solution to these challenges [18]. Furthermore, 3D stacking can incorporate disparate technologies and device types to represent a wide range of neuron types, neural architectures, AI subsystems such as sensor and communication components in a fully integrated systems framework to achieve better performing AI solutions and AGI [27][40]. Recent studies have shown that in some use cases it is possible to directly train the neuromorphic chips through external sensory data, which can fundamentally change AI training through neuromorphic architectures.

*2.1.4* **Memory Bottleneck & GPU Cache Stacking for Transformers:** Transformer architectures have found a wide range of applications in recent years from question answering and AI assistants to financial crime detection and time series predictions. Current AI training is predominantly performed on graphical processing units GPUs along with some TPUs. However, GPUs' limited on-chip cache and memory capacity requires extensive optimizations at the algorithmic level such as grouped and multi query attention mechanisms for transformers trading off small amounts of representation accuracy with memory optimization [55][57]. In response, 3D integrating cache layers with GPUs have been proposed and is pursued for products [56]. This highlights the need to analyze the power, thermal, noise and reliability implications of stacking high activity layers such as GPUs with high sensitivity layers such as caches, memory and non-volatile memories through hardware characterization.

## 2.2 Full System Integration through 3D
Even outside of the rapidly growing AI hardware and systems market, 3D integration provides numerous advantages to mainstream computational systems ranging from mobile processors to high-end servers and supercomputers. One such advantage is the ability to integrate system components that range from processor cores and accelerators to non-volatile memories, DRAM and eDRAM layers in a coherent and fully integrated stack. This provides key benefits such as improved performance (due to reduced footprints and clocking speeds), improved communication efficiency by eliminating off-chip communication bottlenecks, and performance advantages from bringing data to processing units.

## 2.3 Multi-layer Processor & Cache Stacking
In recent years, data-centric computing has been seen as a potential solution to eliminate the memory bottlenecks of von Neumann architectures. 3D integration of multiple processing and cache/memory layers have emerged as an important enabler to improve performance for data centric computing [44] [46]. Processor stacks with memory integration have been shown to provide significant performance advantages [22][25] yet suffer from similar challenges with GPU/cache stacks.

## 2.4 Cost, Heterogeneity and Modularity Benefits
Another, less discussed, advantage of 3D integration is the ability to build customized systems in a modular fashion at a low cost. By introducing standardized layer interfaces and power delivery structures, a large number of custom device layers to achieve a wide range of custom products and system configurations as well as achieving higher system performance. It allows for the manufacturing of individual device layers at different generations (such as older generation nodes) at a lower cost. The enhanced modularity and cost advantages serve as important 3D drivers in commercial use cases [26] [42].

## 2.5 System Yield & Reliability Improvements
The underlying modularity and cost effectiveness of 3D systems also provide the opportunity to introduce redundant architectural blocks (such as cores, caches, accelerators, communication macros) for improved lifetime reliability and yield as well as higher performance. In some use cases, 3D chiplets with redundant blocks were used to repair faulty chip components as well [54].

## 2.6 Recent 3D Solutions
A number of 3D processor systems have been presented in recent years. Cache stacks attach additional SRAM caches to the design [11]. Similarly, systems that use power-delivery chip 3D integrated with its AI processor using wafer-on-wafer bonding technology have been developed [13]. Such systems provide improved performance and energy efficiency through the enhanced power delivery infrastructure. Other systems use 2.5D integration to integrate multiple chiplets as well as processor and cache stacking [12] using compute and cache tiles as well as heat removal tiles through the integration with the base tile. 3D integration of neuromorphic chip solutions has been announced by some chip makers[14]. More recently, 3D stack of computing, memory, and communication components for 3 layers processing tens of terabytes of data vertically between the individual device layers have been demonstrated for supercomputers [41].

## 3. Need for 3D Characterization & Emulation
Despite the bespoke advantages, trends and drivers, heterogeneous 3D stack design is a complex process plagued with numerous challenges. In heterogeneous 3D stacks, the number of system components, their characteristics and the number of interactions among the components increase rapidly. These include block activity levels, power, thermal, noise characteristics, soft error vulnerabilities, error rates and lifetime reliability profiles. The process is further complicated by the complex material composition of 3D designs with multiple, some thinned, silicon layers, back-end-of-line layers of various thicknesses, dielectrics, micro-C4 and TSV structures. This causes large, non-convex and complex design spaces that are hard to optimize for, both during design time and for run-time management. Yet, current modeling frameworks lag behind 3D manufacturing capabilities. Moreover, as the complexity of the 3D systems and their number of components grow, the existing modeling frameworks such as multi-physics solvers exhibit more severe limitations. Long term characterization such as lifetime reliability modeling further complicates the process as it doesn't lend itself well to modeling frameworks. Yet, the design of 3D systems necessitates a comprehensive understanding of design implications on various components and the overall system. In this study, we introduce a configurable hardware emulation vehicle design that facilitates real-time characterization of a broad spectrum of stacking alternatives with fine grain analysis capabilities, multi-dimensional analysis of various criteria and their interactions as well as long-term analysis for lifetime reliability.

### 3.1. Architectural Overview

The 3D emulator chiplet design is geared towards fundamental insights into 3D technology and encompasses the following sub-systems:

*(i) On-Chip Sensor Infrastructure:* This infrastructure enables the thermal and performance characterization of multi-layer stacks. A controller unit governs the on-stack sensors, orchestrating their interactions with on-chip actuation and sensor structures. As shown in earlier studies the placement of thermal sensors is important in properly identifying run-time heating characteristics. As a result, the design includes a large number of sensors to capture the detailed heating profile [23].

*(ii) Configurable Heat Generator Structures:* The tiled 3D processor emulator design allows for precise adjustments of heat and noise levels within each stratum and tile. This is made possible through configurable thermal macros. The on-chip controller unit for each stratum/tile manages the desired noise and heat levels. The resulting tiles function as core proxies, capable of replicating a wide range of processor, GPU and accelerator power, noise characteristics, cache, and memory layers. Activating and tuning the power/heat levels in individual heat generator tiles can produce a variety of lateral and vertical thermal gradients.

*(iii) Modular Embedded Tile Infrastructure:* This infrastructure is designed to incorporate built-in sensing and characterization capabilities within customized stacking alternatives. The embedded design can be integrated into various stack configurations as sensor and core proxy blocks. The design also reflects the detailed thermal models used to capture the vertical heat flow challenges in thinned device layers. As discussed in [24] dense vertical TSVs such as bus blocks may serve as thermal blockages that increase thermomechanical stress and resulting reliability issues. The design planning for this study involved simulation-based analysis of temperature, encompassing considerations such as decoupling capacitor placement and temperature characterization [2].

### 3.2. 3D Stack Design

The emulator was designed and manufactured using 45nm SOI base technology, incorporating new features to enable 3D stacking [1], [2], [21], [36] as discussed in [21]. Key highlights of these new 3D features encompass: (i) Through silicon vias (Copper TSVs) (ii) Thinned silicon chip layers (of 50 μm thickness). (iii) Backside redistribution layers on the reverse side of thinned silicon wafers. (iv) MicroC4 (μC4) contact bumps. 4-layer stack configuration is shown in Figure 1. A standard 45nm C4 bump, each with approximately 80 μm in diameter, connects the package to the thinned chip. On top of the C4 contacts shows the standard metalization (back-end-of-the-line) BEOL layers. The process involves linking the copper through silicon vias from the mid-level BEOL through the device layers and continues through the silicon wafer. At the back side of the wafer, through silicon vias connected through the redistribution layers and connected to the micro-C4s. The top layer chip is devoid of TSVs and is cooled through a water cooled heat sink solution.

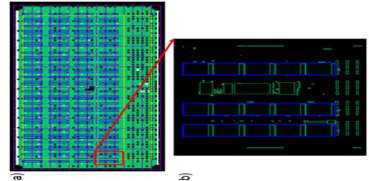

Figure 2 displays the floorplan of the emulator chip, as follows: a) floorplan image of the emulator chip. b) A magnified view of the floorplan, offering a closer look at the macros. Each blue column houses programmable power generators, while thermal sensors are positioned between the second and third blue columns. Additional elements encompass control logic, inter-chip IO, and buffers within other boxes [21].

The emulator chip has an approximate size of *12mm x 6mm* with configurability of stacking one to four layers of chips as discussed by Kursun et.al.[20] and [21]. Three distinct mask sets for *S0, SN(x),* and *SP* were designed for the different layers, based on their exact position in the entire stack. For the different configurations the layer orders are as follows: 2L Stack *(SP, S0 and Heat Sink)*, 3L Stack *(SP, SN, S0, Heat sink),* 4L Stack *(SP, SN2, SN1, SO, Heat Sink).* In each of these configurations the SP chips are connected to the package through the C4 contacts and the SP chips are equipped with TSVs to integrate with the other layers in the stack. The top layer S0 chip is in contact with the heat sink with *300-750 um* thickness. Due to its position, this layer does not include any TSVs. The mid-layers in the multi-layer stacks were designed with BEOL layers to connect to uC4, TSV, FEOL, backside metal and uC4s. Emulator tiles consist of activity generators and sensors [21]. Fig 1 shows columns of power generators, the uC4s as well as probe pads for wafer processes. Below the array of configurable power generator columns in the floorplan are the PLL, IO's, boundary scan, and global control, which is shown in more detail in Figure 1b. The thermal sensor infrastructure was placed to maximize the thermal profile extraction from the multitude of power generators in a number of different configurations.

### 3.3 Thermal Modeling

Multi-layer chips are challenged by thermal issues due to the underlying complexity of the thermal interactions among layers and their power consumption, as well as the complexity of the structure of the 3D stack with multitude of silicon and BEOL layers, vias, insulators and other structures. Having an accurate thermal modeling and prediction of the 3D stack is critically important, as the 3D designs are frequently constrained by the thermal behaviors and cooling challenges. The thinned device layers *(SP and SNx)* away from the heat sink are even more challenging from a thermal design perspective as they face the complex structure of the 3D stack for heat dissipation as well as being further away from the heat sink (as shown in Figure 2). This results in each layer from the heat sink being higher in temperature than the ones before, despite having similar power dissipation and power density values. Experimental analysis using *ANSYS* and *Flotherm CFD* thermal analysis infrastructure of the multi-layer stack indicates temperature difference between the S0 stratum and the thinned device layers in some configurations. These results and the significant thermal differences were confirmed by 3D testing and thermal measurements of the stack in [21] [43].

Figure 3 shows the thermal map of the S0 layer of the multi-layer emulator stack. There are fundamental differences in the heat dissipation patterns between the *S0* layer and the thinned device layers *SNx*. Detailed thermal analysis and modeling of the 3D stacking options have been completed for experiment design [30] [31] [32] [48]. The complex interactions among macros *(top, down, lateral)* from multiple device layers often require targeted thermal analysis and temperature-aware design planning for 3D systems. Furthermore, 3D hotspots, especially in thinned device layers can cause significant mechanical stress, cracking and reliability issues.Such challenges exacerbate when temperature-aware design planning is not used for all design stages (including but not limited to placement, routing and other physical design stages). As an example, the placement of densely populated TSV regions with hotspot proximity has been shown to create reliability issues [34].

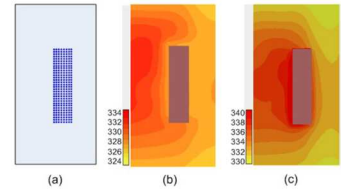

Figure 2. Thermal maps of alternative TSV-farm structures in a thinned 10μm layer. (a) Top view of the dense via farm (b) Highly conductive copper TSV implementation; (c) Thermal blockage due to the composite, insulatedTungsten via farm, with 6 degrees of temperature increases. [35]

As discussed before, emulator infrastructure provides the ability to mimic the characteristics of a wide range of components such as processor layers or cores, GPUs, accelerators, logic, memory and non-volatile memory layers in great detail due to the fine-grain composition of the noise generators [45][36]. The corresponding thermal sensor infrastructure captures the fine-grain changes in the temperatures of the individual device layers based on the activity generation. The use of noise generators and dynamic configuration of thermal hotspot behavior with fine grain control over the temporal characteristics provides the opportunity to analyze thermomechanical stress emulation for complex processor stacking architectures with local heating behavior. This profiling is further used for fine-tuning not only thermal but also mechanical models for accurate representation and analysis of future stacking options. As a result, a diverse set of run-time thermal management techniques can be implemented using the described emulator [33] [47] [50].

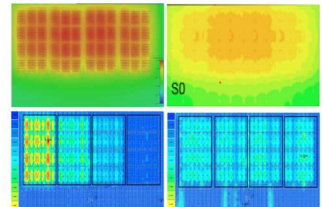

Figure 3. Thermal profiling of the 4-layer stack architecture device layers through the use of noise generators [21]

### 3.4 Noise Generation for 3D System Emulation

The vertical placement of various functional blocks through power delivery columns can create challenges in 3D systems [53] [54]. The system also provides the opportunity to also analyze power delivery and power supply noise for a wide range of scenarios [29]. Noise analysis of the chip provides important insights for numerous stacking options such as SRAM, DRAM, eDRAM and NVM stacking, processor, logic, GPU and accelerator stacking as well as neuromorphic computing chips. The emulator infrastructure provides the capability to analyze a spectrum of scenarios such as noisy processor layers being integrated with sensitive device layers and storage. In addition to emulating various stacking scenarios, noise generators provide further hardware security benefits to 3D chips which is demonstrated by the emulator framework as well. By largely eliminating side-channel information leakage from 3D chips they improve the overall security of the 3D system [28].

## 4. Conclusions

3D system integration not only provides an alternative to scaling laws, it enables performance gains by improving the interconnectivity between device layers, largely eliminating slow and inefficient off-chip communications, reducing memory bottleneck challenges and providing heterogeneous integration of the full system. Moreover, in recent years, AIs unprecedented computational demands combined with its underlying energy inefficiency have increased the interest in 3D systems. Achieving low-power, energy efficient, reliable and robust 3D systems, including custom 3D AI hardware stacks, requires detailed understanding of the underlying power, thermal, noise, soft-error and lifetime reliability characteristics of the system components. This is especially important when system components are heterogeneous and engage in complex multi-dimensional interactions on a heterogeneous substrate, where individual components may further have different sensitivities to power, thermal, noise, reliability and other factors. The 3D emulator infrastructure provides a deeper understanding of the characteristics of 3D systems by fine-grain analysis of its components, as well as their complex interactions through a flexible and configurable actuation and sensing infrastructure. These insights are then used for 3D system design stages from placement to routing and thermal sensor placement, calibrating the corresponding performance, temperature, power, mechanical and reliability models and developing effective run-time management techniques.